\documentstyle[prl,aps]{revtex} 
\twocolumn

\begin{document}
\author{Jian Qi Shen $^{1,}$$^{2}$} 
\address{$^{1}$  Centre for Optical
and Electromagnetic Research, State Key Laboratory of Modern
Optical Instrumentation, Zhejiang University,
Hangzhou SpringJade 310027, P.R. China\\
$^{2}$Zhejiang Institute of Modern Physics and Department of
Physics, Zhejiang University, Hangzhou 310027, P.R. China}
\date{\today }
\title{An Experimental Realization of Quantum-vacuum Geometric Phases \\by Using the Gyrotropic-medium Optical Fiber}
\maketitle
\begin{abstract}
{\it Abstract} 

The connection between the {\it quantum-vacuum geometric phases}
(which originates from the vacuum zero-point electromagnetic
fluctuation) and the non-normal product procedure is considered in the
present Letter. In order to investigate this physically
interesting geometric phases at quantum-vacuum level, we suggest an
experimentally feasible scheme to test it by means of a noncoplanarly
curved fiber made of gyrotropic media. A remarkable feature of the present experimental realization is that
one can easily extract the nonvanishing and nontrivial quantum-vacuum geometric
phases of left- and/or right- handed circularly polarized light from
the vanishing and trivial total quantum-vacuum geometric phases.
\\ \\ 

\end{abstract}
Since Berry discovered that a topological (geometric) phase exists
in quantum mechanical wavefunction of time-dependent systems,
geometric phase problems have captured considerable attention of
researchers in various fields such as quantum mechanics\cite{Berry},
differential geometry\cite{Simon}, gravity theory\cite
{Furtado,Shen1}, atomic and molecular physics\cite
{Kuppermann,Kuppermann2,Levi}, nuclear physics\cite {Wagh},
quantum optics\cite{Gong}, condensed matter
physics\cite{Taguchi,Falci}, molecular systems and chemical
reaction\cite {Kuppermann} as well. More recently, many authors
concentrated their particular attention on the potential applications of
geometric phases to the geometric quantum computation, quantum
decoherence and related topics\cite{Taguchi,Wu,Jones,Wang,Zhusl}. One of the most important physical
realizations of Berry's phase ({\it i.e.}, cyclic adiabatic
geometric phase) is the model describing the propagation of photons inside a helically
curved optical fiber, which was proposed by Chiao and Wu\cite{Chiao}, and later
performed experimentally by Tomita and Chiao\cite{Tomita}. Afterwards, a large
number of investigators treated this photon geometric phases by
making use of the classical Maxwell's electrodynamics,
differential geometry method (parallel transport) and quantum
adiabatic theory both theoretically and experimentally\cite{Kwiat,Robinson,Haldane1,Haldane2}. Based on
the above investigations, we studied the nonadiabatic noncyclic
geometric phases of photons propagating inside a noncoplanarly
curved optical fiber\cite{Zhu,Gao2} by means of the Lewis-Riesenfeld invariant
theory\cite{Riesenfeld} and the invariant-related unitary transformation
formulation\cite{Gao1}. By using the obtained results\cite{Zhu,Gao2}, we considered the photon helicity inversion in
the curved fiber and its potential applications to information
science\cite{Shenpla} and proposed a second-quantized spin model to describe the coiled light in a curved fiber, where
the vacuum zero-point fluctuation is involved. As was stated by Fuentes-Guridi {\it et al.}, 
in a strict sense, the Berry phase has been studied only in a semiclassical 
context until now\cite{Fuentes}. Thus the effects of the vacuum field on the 
geometric evolution are still unknown\cite{Fuentes}.  In their paper\cite{Fuentes},
Fuentes-Guridi {\it et al.} considered the time evolution of a
spin-$1/2$ particle interacting with a second-quantized external
magnetic field and proposed a vacuum-induced spin-$1/2$ Berry's
phase, which they regarded as the effect of vacuum photon
fluctuation. In this
Letter, we will propose a new nontrivial vacuum effect, {\it i.e.}, 
the quantum-vacuum geometric phases, and study its novel properties
(particularly its connection with normal-order procedure in
quantum field theory), and then suggest an experimental realization
of this geometric phases at quantum-vacuum level by using the
gyrotropic-medium fiber. 

Note that here the quantum-vacuum geometric phases of photons
results from the zero-point energy of vacuum quantum fluctuation.
This, therefore, means that this geometric phases is quantal in
character and, moreover, has no classical counterpart, namely, it
cannot survive the correspondence-principle limit into the
classical level. It is well known that in the conventional quantum
field theory, both infinite vacuum fluctuation energy and
divergent vacuum electric charge density are removed by the so-called
{\it normal-order procedure} and new vacuum backgrounds of quantum
fields, in which the vacuum expectation values of both charge
density and Hamiltonian vanish, are therefore re-defined. Since in
the {\it time-independent} quantum field theory, the infinite
constant is harmless and easily removed, the normal-order
procedure applied to these {\it time-independent} cases is
reliable and valid indeed. However, in the {\it time-dependent} quantum
field theory (such as quantum field theory in curved space-time,
{\it e.g.}, time-dependent gravitational backgrounds and expanding
universe), the time-dependent vacuum zero-point fields may also
participate in the time evolution process and therefore cannot be
regarded merely as an inactive onlooker. We think that in these
{\it time-dependent} cases, the non-normal order for operator product in
quantum field theory should be taken into account and one can
therefore predict some physically interesting quantum-vacuum
effects. In what follows by analyzing the time evolution of
photon wavefunction we consider such vacuum effect in
a time-dependent quantum system, {\it i.e.}, the Chiao-Wu case\cite{Chiao,Tomita,Zhu} 
where the rotation of photon polarization planes, which gives rise to photon geometric phases, in a
noncoplanar fiber takes place.  
   
According to the Liouville-Von Neumann equation\cite{Zhu}, 
where the Lewis-Riesenfeld invariant\cite{Riesenfeld} is the photon helicity $h={\bf k}\cdot {\bf{S}}/k$, 
the effective Hamiltonian that describes
the propagation of coiled light in a curved fiber is of the form\cite{Zhu,Gao2}   
$H_{\rm eff}(t)=[{\bf{k}}(t)\times
\dot{\bf{k}}(t)]\cdot {\bf{S}}/k^{2} $
with ${\bf k}$, $k$ and ${ \bf{S}}$ representing the wave vector, the magnitude 
of ${\bf k}$ and the spin operator of photons. The dot denotes the derivative of ${\bf k}$ with respect to time. 
Thus the time-dependent Schr\"{o}dinger equation governing the time evolution of 
photon wavefunction inside the noncoplanarly curved fiber is written in the form\cite{Zhu,Gao2}  (in the unit $\hbar=c=1$) 
\begin{equation}
i\frac{\partial \left| \sigma ,{\bf{k}}(t)\right\rangle }{\partial t}=\frac{%
{\bf{k}}(t)\times \dot{\bf{k}}(t)}{k^{2}}\cdot {\bf{S}}\left|
\sigma ,{\bf{k}}(t)\right\rangle,         \label{eq1}
\end{equation}
where $\sigma=\pm 1$ is the eigenvalues of photon helicity $h$ corresponding to the           
right- and left- handed circularly polarized photons. The photon wave vector ${\bf k}$ inside the fiber
can be written in the spherical polar coordinate system, {\it i.e.}, 
${\bf k}=k(\sin\theta\cos\varphi, \sin\theta\sin\varphi, \cos\theta)$, which is
always along the tangent to the curved fiber at each point at
arbitrary time. By making use of the Lewis-Riesenfeld invariant
theory and the invariant-related unitary transformation
formulation\cite{Riesenfeld,Gao1}, we obtain the exact solutions $
\left| \sigma ,{\bf{k}}(t)\right\rangle=\exp \left[\frac{1}{i}\phi
_{\sigma }^{\rm (g)}(t)\right]V(t)\left| \sigma,k \right\rangle$ 
to Eq.(\ref{eq1}), where $\left| \sigma,k \right\rangle\equiv\left|
\sigma,{\bf{k}}(t=0) \right\rangle$ is the initial photon
polarized state, $V(t)=\exp[\beta(t){\bf S_{+}}-\beta^{\ast}(t){\bf S_{-}}]$ 
with $\beta(t)=-[\theta(t)/2]\exp[-i\varphi(t)]$, $\beta^{\ast}(t)=-[\theta(t)/2]\exp[i\varphi(t)]$\cite{Zhu} 
and ${\bf S}_{\pm}={\bf S}_{1}\pm i{\bf S}_{2}$. The noncyclic nonadiabatic geometric phase is given as follows

\begin{equation}
\phi _{\sigma }^{\rm
(g)}(t)=\left\{{\int_{0}^{t}\dot{\varphi}(t^{^{\prime }})[1-\cos
\theta (t^{^{\prime }})]{\rm d}t^{^{\prime
}}}\right\}\left\langle \sigma,k \right| S_{3}\left| \sigma,k
\right\rangle .            \label{eq4}
\end{equation}
It is apparent that in the adiabatic cyclic case for the Chiao-Wu's coiled light in a helically curved fiber, where $\dot{\theta}=0$ 
and $\dot{\varphi}$ (expressed by $\Omega$ that is constant) is the rotating frequency of photon moving on the fiber helicoid, 
the geometric phase in a cycle ($T=2\pi/\Omega$) over the photon momentum space takes the form 
$\phi _{\sigma }^{\rm (g)}(T)=2\pi(1-\cos \theta )\left\langle
\sigma,k \right| S_{3}\left| \sigma,k \right\rangle$,
where the expression $2\pi(1-\cos \theta )$ denotes the solid angle subtended by a curve traced by the wave vector at the center of photon
momentum space. This fact demonstrates the topological and global properties of geometric phases. This shows that the above calculation is self-consistent.

Now we consider the expectation value, $\left\langle
\sigma,k \right| S_{3}\left| \sigma,k \right\rangle$, of the third component of photon spin 
operator in Eq.(\ref{eq4}). 
Substitution of the Fourier expansion series of three-dimensional magnetic vector
potentials ${\bf
A({\bf x},t)}$ into the expression $S_{ij}=-\int{(\dot{A}_{i}A_{j}-\dot{A}_{j}A_{i})}{\rm d}^{3}{\bf
x}$ for the spin operator of photon fields yields\cite{Zhu} 
\begin{eqnarray}
S_{3}=\frac{i}{2}[a(k,1)a^{\dagger}(k,2)-a^{\dagger}(k,1)a(k,2)   \nonumber \\
-a(k,2)a^{\dagger}(k,1)+a^{\dagger}(k,2)a(k,1)]
\label{eq6}
\end{eqnarray}
with $a^{\dagger}(k,\lambda)$ and $a(k,\lambda)$ ($\lambda=1, 2$) being the creation and annihilation operators of polarized
photons corresponding to the two mutually perpendicular real unit polarization vectors. Note that here $S_{3}$ is of the non-normal-order form.

In what follows we define the creation and annihilation operators,
$a_{R}^{\dagger}(k)$, $a_{L}^{\dagger}(k)$, $a_{R}(k)$,
$a_{L}(k)$, of right- and left- handed circularly polarized light
\cite{Bjorken}, which are expressed in terms of $a^{\dagger}(k,\lambda)$ and $a(k,\lambda)$, {\it i.e.},        
$a_{R}^{\dagger}(k)=1/\sqrt{2}[a^{\dagger}(k,1)+ia^{\dagger}(k,2)]$,
$a_{R}(k)=1/\sqrt{2}[a(k,1)-ia(k,2)]$,           
$a_{L}^{\dagger}(k)=1/\sqrt{2}[a^{\dagger}(k,1)-ia^{\dagger}(k,2)]$ and
$a_{L}(k)=1/\sqrt{2}[a(k,1)+ia(k,2)]$. Thus Eq.(\ref{eq6}) can be rewritten in 
terms of the creation and annihilation operators of right- and left- handed polarized photons, namely,
\begin{eqnarray}
S_{3}=\frac{1}{2}[a_{R}(k)a_{R}^{\dagger}(k)+a_{R}^{\dagger}(k)a_{R}(k)]   \nonumber \\
-\frac{1}{2}[a_{L}(k)a_{L}^{\dagger}(k)+a_{L}^{\dagger}(k)a_{L}(k)],
\label{eq7}
\end{eqnarray}
which can also be rewritten as $S_{3}=[a_{R}^{\dagger}(k)a_{R}(k)+1/2]-[a_{L}^{\dagger}(k)a_{L}(k)+1/2]$.

The monomode multiphoton states of left- and right- handed (LRH)
circularly polarized light (at $t=0$) can be defined as
$|\sigma=-1,k,n_{L}\rangle=(n_{L}!)^{-1/2}[a_{L}^{\dagger}(k)]^{n_{L}}|0_{L}\rangle$ and
 $|\sigma=+1,k,n_{R}\rangle=(n_{R}!)^{-1/2}[a_{R}^{\dagger}(k)]^{n_{R}}|0_{R}\rangle$
with $n_{L}$ and $n_{R}$ being the LRH polarized photon occupation
numbers, respectively. In the following we will calculate the geometric
phases of multiphoton states $
|\sigma=+1,k, n_{R}; \sigma=-1,k, n_{L}\rangle$, which is the direct product of LRH multiphoton states, {\it i.e.}, $|\sigma=+1,k,
n_{R}\rangle\otimes|\sigma=-1,k, n_{L}\rangle$. Insertion of the expression for the monomode multiphoton polarized states into Eq.(\ref{eq4})
yields the geometric phases of multiphoton polarized states, $\phi^{\rm (g)}(t)=(n_{R}-n_{L})\left\{{\int_{0}^{t}\dot{\varphi}(t^{^{\prime
}})[1-\cos \theta (t^{^{\prime }})]{\rm d}t^{^{\prime
}}}\right\}$, which is independent of $k$ but dependent on the geometric
nature of the pathway (expressed in terms of $\theta$ and
$\varphi$) along which the light wave propagates. Although the
phases $\phi^{\rm (g)}(t)$ associated with the
photonic occupation numbers $n_{R}$ and $n_{L}$ are quantal geometric
phases of photons\cite{Gao2}, they do not belong to the geometric phases at
quantum-vacuum level which arises, however, from the zero-point
electromagnetic energy of vacuum quantum fluctuation.

It follows from the expression (\ref{eq7}) for $S_{3}$ that both the
geometric phases of left- and right- handed circularly polarized
photon states, {\it i.e.}, $|\sigma=-1,k, n_{L}\rangle$ and
$|\sigma=+1,k, n_{R}\rangle$, are respectively as follows
\begin{eqnarray}
 \phi_{L}^{\rm (g)}(t)=-(n_{L}+\frac{1}{2})\left\{{\int_{0}^{t}\dot{\varphi}(t^{^{\prime
}})[1-\cos \theta (t^{^{\prime }})]{\rm d}t^{^{\prime
}}}\right\},  \nonumber \\             
 \phi_{R}^{\rm (g)}(t)=+(n_{R}+\frac{1}{2})\left\{{\int_{0}^{t}\dot{\varphi}(t^{^{\prime
}})[1-\cos \theta (t^{^{\prime }})]{\rm d}t^{^{\prime
}}}\right\}.                 \label{eq11}
\end{eqnarray}

According to Eq.(\ref{eq11}), it is readily verified that the time-dependent zero-point energy possesses
physical meanings and therefore contributes to geometric phases of
photon fields. Thus the noncyclic nonadiabatic geometric phases
of left- and right- handed polarized states at quantum-vacuum
level are of the form
\begin{equation}
 \phi_{\sigma=\pm 1}^{\rm (vac)}(t)=\pm\frac{1}{2}\left\{{\int_{0}^{t}\dot{\varphi}(t^{^{\prime
}})[1-\cos \theta (t^{^{\prime }})]{\rm d}t^{^{\prime
}}}\right\}.                       \label{eq12}
\end{equation}   

However, it should be pointed out that, unfortunately, even at the
quantum level, this quantum-vacuum geometric phases
$\phi_{\sigma=\pm 1}^{\rm (vac)}(t)$ that is observable in principle is absent in the previous fiber
experiments\cite{Tomita,Kwiat,Robinson,Haldane1}, since it follows from (\ref{eq11}) and (\ref{eq12})
that the signs of quantal geometric phases of left- and
right- handed circularly polarized photons are just opposite to one
another, and so that their quantum-vacuum geometric phases (\ref{eq12}) are
counteracted by each other. Hence the observed geometric phases
are only those associated with the creation operators $a^{\dagger}_{L}$ and $a^{\dagger}_{R}$ of LRH polarized photons, the cyclic adiabatic case of which 
has been measured in the optical fiber experiments
performed by Tomita and Chiao {\it et
al.}\cite{Tomita,Kwiat,Robinson,Haldane1}. Although the total of LRH quantum-vacuum geometric phases (\ref{eq12}) is trivial, 
the vacuum geometric phase of separate circularly polarized field is nontrivial, which deserves investigation 
experimentally. The troublesome problem left to us now is that how can we detect the above quantum-vacuum geometric phases 
of left- and/or right- handed polarized fields that has been cancelled by each other?      

More recently, we suggest a new scheme to test the existence of this vacuum effect, the 
idea of which is to extract the nonvanishing cyclic quantum-vacuum geometric phases $\phi_{\sigma=+1}^{\rm (vac)}(T)$ or $\phi_{\sigma=-1}^{\rm (vac)}(T)$ by changing 
the mode distribution structures of vacuum photon field (or inhibiting vacuum photon fluctuation of 
certain propagation constant). 
This is not strange to us. For example, it
is well known that in Casimir's effect the vacuum-fluctuation
electromagnetic field in a finitely large space ({\it i.e.}, the
space between two parallel metallic plates) will alter its mode structures, namely, the
zero-point field with wave vector $k$ less than
$\sim\left(\frac{\pi}{a}\right)$ does not exist in this space with
a finite scale length $a$. Another illustrative example is the inhibition of spontaneous emission in photonic crystals\cite{Yablonovitch} and cavity resonator\cite{Hulet}.

For this aim, we take into account the peculiar wave propagation inside a kind of anisotropic materials (gyrotropic media), the 
electric permittivity and magnetic
permeability of which are tensors taking the following form\cite{Veselago} 
\begin{equation}
\hat{\epsilon}=\left(\begin{array}{cccc}
\epsilon_{1}  & i\epsilon_{2} & 0 \\
-i\epsilon_{2} &   \epsilon_{1} & 0  \\
 0 &  0 &  \epsilon_{3}
 \end{array}
 \right),                 \quad          \hat{\mu}=\left(\begin{array}{cccc}
\mu_{1}  & i\mu_{2} & 0 \\
-i\mu_{2} &   \mu_{1} & 0  \\
 0 &  0 &  \mu_{3}
 \end{array}
 \right)              
 \label{eq13}
\end{equation}
Assuming that the direction of the electromagnetic wave vector
${\bf k}$ is parallel to the third component of the Cartesian
coordinate system, with the help of Maxwell's Equations, one can arrive at
\begin{equation}
 n_{\pm}^{2}=(\epsilon_{1}\pm \epsilon_{2})(\mu_{1}\pm \mu_{2}),               
 \label{eq14}
\end{equation}
where $n_{+}$ and $n_{-}$ are the optical refractive indices squared of such gyrotropic media 
corresponding to the right- and left- handed circularly polarized light, respectively\cite{Veselago,arxiv}. Since in such gyrotropic media, one is
positive and the other negative for the optical refractive index
squared $n^{2}$ corresponding to the two directions of
polarization of the electromagnetic wave, only one wave can
propagate in gyrotropic media. So, the quantum-vacuum geometric
phases of LRH polarized photons cannot be
eliminated by each other and it is therefore possible for physicists 
to easily test the remainder of quantum-vacuum geometric phases experimentally. 
If, for example, by taking some
certain values of $\epsilon_{1}$, $\epsilon_{2}$, $\mu_{1}$ and
$\mu_{2}$, then $n_{L}^{2}<0$ while $n_{R}^{2}>0$ and consequently
the left-handed polarized light cannot be propagated in this
medium, and in the meanwhile the quantum vacuum fluctuation
corresponding to the left-handed polarized light will also be
inhibited ({\it e.g.}, the wave amplitude exponentially decreases
because of the imaginary part of the refractive index $n_{L}$) in this
anisotropic absorptive medium. Thus the vacuum-fluctuation
electromagnetic field alters its mode structures in the absorptive
medium. For this reason, the only retained geometric phases is
that of right-handed polarized light, which we can detect
experimentally. 

As an illustrative example, we now discuss the light propagation inside an
optical fiber made of gyrotropic media. We only consider the
condition under which
$|\epsilon_{2}|>|\epsilon_{1}|=-\epsilon_{1}$ and $\mu_{1}\pm \mu_{2}>0$. If, for
instance, $\epsilon_{2}$ is positive, then
the right-handed polarized light can be propagated
while the left-handed polarized light cannot be
propagated in the fiber (because of the negative $n_{-}^{2}$ and the consequent
imaginary propagation constant $k_{-}$, which is expressed by
$n_{-}{\omega}/{c}$); conversely, if $\epsilon_{2}$ is negative,
then the left-handed polarized light can be propagated
while the right-handed polarized light is inhibited
from being propagated (due to the imaginary propagation constant
$k_{+}$, which equals $n_{+}{\omega}/{c}$). Thus in the former case
the phase $\phi_{R}^{\rm (vac)}(T)$ of right-handed polarized
light, and in the latter case the phase $\phi_{L}^{\rm
(vac)}(T)$ of left-handed polarized light instead, may respectively be
detected in this gyrotropic-medium fiber experiment. 

Since the vacuum photon fluctuation with $k$ less than $\pi/a$ will be inhibited in the 
space between two parallel conducting plates whose separation is $a$, we can suggest another scheme
to detect $ \phi_{\sigma=\pm 1}^{\rm (vac)}(T)$: specifically, if $\epsilon_{1}$ and $\epsilon_{2}$ (or $\mu_{1}$ and $\mu_{2}$) of gyrotropic medium 
are chosen to be $\epsilon_{1}=\epsilon_{2}$ (or $\mu_{1}=\mu_{2}$), then the vacuum fluctuation corresponding to the left-handed polarized light is inhibited since its propagation constant 
$k_{-}$ tends to zero (and hence the wavelength is larger than the space scale $a$). Thus the only retained vacuum geometric phase is that of right-handed polarized light.

The physical significance of the subject presented in this Letter may be as follows:

(i) the quantum-vacuum geometric phases found here possesses interesting properties, for it has an important connection
with the topological nature of time evolution of quantum vacuum fluctuation. To the best of our knowledge, 
in the literature, less attention was paid to such geometric phases at purely quantum-vacuum level. 
According to Fuentes-Guridi's statement, such vacuum geometric phases may open up a new areas to 
the study of the consequences of field quantization in the geometric evolution of states\cite{Fuentes}.
It is emphasized that this vacuum effect deserves detailed consideration both theoretically and experimentally;

(ii) in order to extract the nontrivial quantum-vacuum geometric phases of polarized light from the trivial total quantum-vacuum geometric phases (which has been cancelled by each other and therefore vanishing), a new scheme, which is somewhat ingenious, by using the gyrotropic-medium optical fiber is proposed;

(iii) we think that the detection of quantum-vacuum geometric phases may be essential for 
the investigation of the {\it time-dependent} quantum field theory. As was discussed above, the quantum-vacuum geometric phases is related close to 
the non-normal order for operator product in second quantization. If the existence of quantum-vacuum geometric phases is truly 
demonstrated in experiments, then we should consider the validity problem of normal product technique in the {time-dependent} 
field theory. This, therefore, means that the experimental study of quantum-vacuum geometric phases would be a fundamental 
and important subject.

Recently, many authors applied geometric phases to some areas such as 
quantum decoherence and geometric (topological) quantum computation\cite{Taguchi,Wu,Jones,Wang,Zhusl}. 
It may be believed that the quantum-vacuum geometric phases in the fiber would 
have some possible interesting applications to these subjects. 
We hope all the effects and phenomena presented in this Letter would be investigated
experimentally in the near future.                           
  
\textbf{Acknowledgements}  I thank X.C. Gao for his helpful
proposals. This project was supported in part by the National
Natural Science Foundation of China under the project No.
$90101024$. 
\\ \\
J.-Q. Shen's electronic address: jqshen@coer.zju.edu.cn

\end{document}